\documentclass{iau}
\usepackage{graphicx}

\title[IAUS291.~~Can we see pulsars around Sgr A$^{\bf \star}$?] 
{Can we see pulsars around Sgr A$^{\bf \star}$? \\The latest searches with the Effelsberg telescope. } 

\author[R. P. Eatough \& M. Kramer]  
{R. P. Eatough,
M. Kramer, B. Klein, R. Karuppusamy, D. J. Champion, P. C. C. Freire, N. Wex, \and K. Liu.}

\affiliation{Max-Planck-Institut f\"ur Radioastronomie, Auf dem H\"ugel 69,
53121, Bonn, Germany\\ email: {\tt reatough@mpifr-bonn.mpg.de} }

\pubyear{2012}
\volume{291}  
\jname{\mbox{Neutron Stars and Pulsars: Challenges and Opportunities after 80 years}}
\editors{J. van Leeuwen, ed.} 
\begin{document}

\maketitle

\begin{abstract}
Radio pulsars in relativistic binary systems are unique tools to study
the curved space-time around massive compact objects. The discovery of
a pulsar closely orbiting the super-massive black hole at the centre
of our Galaxy, Sgr A$^{\bf \star}$, would provide a superb test-bed
for gravitational physics. To date, the absence of any radio pulsar
discoveries within a few arc minutes of Sgr A$^{\bf \star}$ has been
explained by one principal factor: extreme scattering of radio waves
caused by inhomogeneities in the ionized component of the interstellar
medium in the central 100 pc around Sgr A$^{\bf \star}$. Scattering,
which causes temporal broadening of pulses, can only be mitigated by
observing at higher frequencies. Here we describe recent searches of
the Galactic centre region performed at a frequency of $18.95$ GHz
with the Effelsberg radio telescope\footnote{\footnotesize This work
  was based on observations with the 100-m telescope of the MPIfR
  (Max-Planck-Institut f\"ur Radioastronomie) at Effelsberg. {\tt
    http://www.mpifr-bonn.mpg.de/effelsberg}}.

\keywords{Pulsars, Sgr A$^{\bf \star}$.}
\end{abstract}


\firstsection 
\section{Introduction}
Both theoretical predictions and observational evidence have shown
that a large population of pulsars should exist in the Galactic centre
(GC) (e.g.  \cite[Pfahl \& Loeb~2004]{pl+04}, \cite[Deneva et
  al. 2009]{den09}, \cite[Wharton et al.~2012]{whar+12}). The
discovery of a pulsar closely orbiting (orbital period $\lesssim$ 1
yr) the supermassive black hole at the centre of our Galaxy, Sgr
A$^{\bf \star}$, would supersede all previous strong field tests of
General Relativity (e.g. \cite[Wex \& Kopeikin 1999]{wk99}, \cite[Liu
  et al. 2012]{liu+12}). Despite many radio pulsar searches that have
covered the GC region, a remarkable lack of pulsars have been
discovered within the central 100 pc. The lack of pulsar detections is
thought to be caused by extreme scattering of radio waves due to
inhomogeneities in the ionized component of the interstellar medium
within 100 pc of Sgr A$^{\bf \star}$ (\cite[Lazio \& Cordes
  1998]{lk98}). Scattering causes temporal broadening of pulses that
has a strong dependence on observing frequency, $\nu$ ($\propto
\nu^{-4}$), rendering pulsar periodicity searches at typical observing
frequencies ineffective e.g. the broadening time is expected to be $\sim\;$500
seconds at 1.4 GHz; 1000 times the length of a typical pulsar period!
Unlike pulse dispersion, which can be corrected for by the use of
filterbanks, scatter broadening cannot be compensated by instrumental
means. As such, pulsar searches covering the GC have been performed at
increasingly high frequencies (e.g. \cite[Johnston et
  al. 1995]{johnston95}, \cite[Kramer et al. 2000]{kram+00},
\cite[Manchester et al. 2001]{mlc+01}, \cite[Klein et
  al. 2004]{klein+04}, \cite[Johnston et al. 2006]{johnston+06},
\cite[Deneva et al. 2009]{den+09}, \cite[Deneva 2010]{den+10},
\cite[Macquart et al. 2010]{macq+10}, \cite[Bates et
  al. 2011]{bates+11}).


\section{Observations \& data analysis}
\begin{figure}[t]
\begin{center}
\includegraphics[width=2.3in, angle=-90]{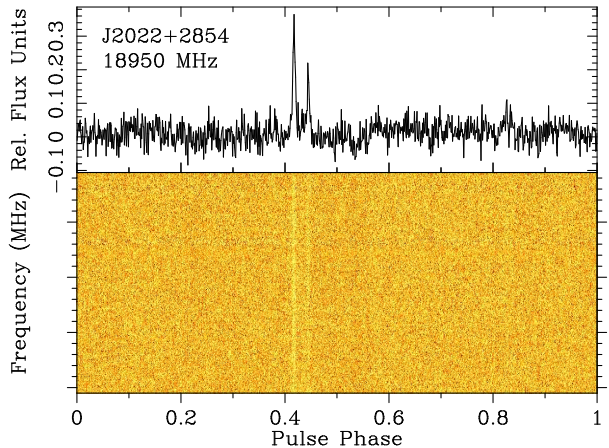} 
\caption{15 minute test observation of PSR B2020$+$28 at a frequency
  of 18.95 GHz using the XFFTS system. The top panel displays the
  integrated profile, and the bottom panel shows phase folded subbands
  across the 2 GHz bandpass.}
\label{fig1}
\end{center}
\end{figure}

From Effelsberg the Sgr A$^{\bf \star}$ region is visible at low
elevation for approximately 2 hours and 25 minutes everyday. For this
reason we have opted to make repeated observations and incoherently
combine the data to improve sensitivity. Observations are made using
the Effelsberg K-band (18-26 GHz) primary focus receiver and the new
XFFTS digital spectrometer operating in pulsar search mode. Using
XFFTS we record dual polarizations for 256 spectral channels across a
bandwidth of 2 GHz. A test observation of PSR B2020$+$28 using our
observing system can be seen in Figure~\ref{fig1}.

\begin{figure}[t]
\begin{center}
\includegraphics[width=3.9in]{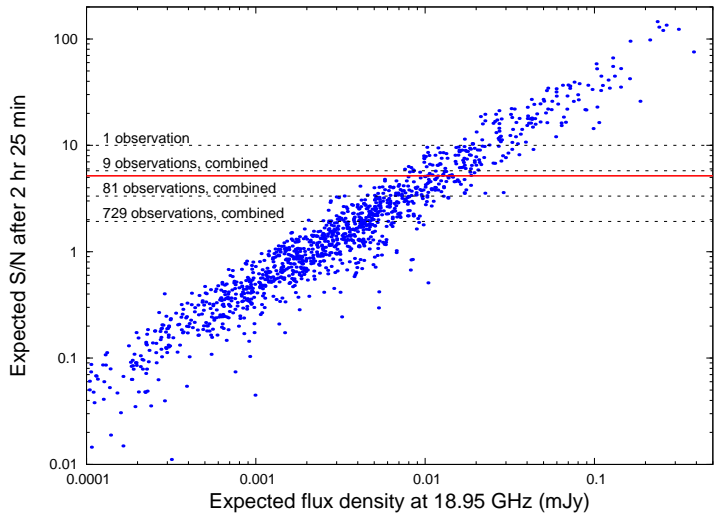}
\caption{Expected detection S/N after a single 18.95 GHz
observation at Effelsberg plotted against the predicted flux densities of
the known pulsar population placed at the distance of the GC. The dotted
horizontal lines indicate the minimum number of combined
observations required to make a detection above our threshold of S/N$\; \sim \;$10.
The red line shows our current sensitivity limit based on a combination of
14 GC observations.}
\label{fig2}
\end{center}
\end{figure}

Data combination is performed by summing the fluctuation spectra of
the dedispersed time series from successive GC observations. So called
`stack searches' have been used to find isolated millisecond pulsars
in the globular cluster Terzan 5 (\cite[Sulman et
  al. 2006]{sulman+06}) and to perform efficient acceleration searches
on the Parkes multi-beam pulsar survey data (\cite[Faulkner
  et~al.~2004]{fsk+04}). We have investigated the sensitivity of a GC
pulsar stack search at Effelsberg by considering the following:
assuming the GC pulsars have the same properties as the current known
population of disk pulsars we have derived their expected flux density
at 18.95 GHz assuming an average spectral index of $-1.8$ and placed
at the distance of the GC. From these values the expected detection
signal-to-noise ratio (S/N) has been calculated for a single 2 hour 25
minute observation. Figure~\ref{fig2} shows the results of this
analysis. From statistical considerations the detection threshold of
this search corresponds to a S/N$\; \sim \;$10. It can be seen that a
number of pulsars would already be possible to detect with a single
observation at 18.95 GHz. The number of pulsars possible to detect
increases as more observations are combined.

\section{Results \& future work}
To date 14 successful observations of the GC region have been
performed. Each observation has been independently searched for
periodic and impulsive signals in addition to a search of the combination of
all 14 observations. In the data taken so far no pulsars have been
detected. Assuming minimal atmospheric contributions to the system
temperature we can place upper limits on the flux density of normal
pulsars (10 per cent duty cycle) within 1.7 pc of Sgr A$^{\bf \star}$
at the observing frequency of 18.95 GHz:
\vspace{0.1cm}
\begin{center}
S/N$\; \sim \;$10 flux density limit of individual observations: $\,\sim 24 \;\mu{\rm Jy}$ \\
S/N$\; \sim \;$10 flux density limit of 14 stacked observations: $\sim 12 \;\mu{\rm Jy}$\\
\end{center}
\vspace{0.15cm} 
We propose to continue observations of the Sgr A$^{\bf \star}$ region
since the emission beams of pulsars in close orbits could precess into
the line-of-sight on the timescale of years (e.g. \cite[Macquart et
  al. 2010]{macq+10}). For the target pulsars our stacking algorithm
poses a problem. Such pulsars may have significant line-of-sight
acceleration as they move in the gravitational potential of Sgr
A$^{\bf \star}$ and thus would show Doppler changes in their pulse
frequency. After data combination, Doppler shifts could smear the
signal and reduce the spectral S/N. To counteract Doppler effects
caused by orbital motion we are investigating `stack-slide searches'
where the spectra are added with different trial frequency offsets.

\end{document}